\documentstyle[11pt]{article}\def\baselinestretch{1.2}
\def\thefootnote{\fnsymbol{footnote}}
\def\bs{\indent \indent}
\def\noin{\noindent}
\def\nonum{\nonumber}
\def\le{\left}
\def\ri{\right}
\def\Tr{{\,\rm Tr}}
\def\Bbar{{\overline {B}}}
\def\pivec{{\vec \pi}}
\def\tvec{{\vec t}}
\def\tauvec{{\vec \tau}}
\def\phihat{{\hat \phi}}
\def\dmu{{\partial_\mu}}
\def\dnu{{\partial_\nu}}
\def\dmup{{\partial^\mu}}
\def\Dmu{{D_\mu}}
\def\del{\partial}
\def\L{{\cal L}}
\def\Lnor{{\cal L}_{\mbox{normal}}}
\def\Lkin{{\cal L}_{\mbox{kin}}}
\def\Lan{{\cal L}_{\mbox{an}}}
\def\Lsu2{{\cal L}_{\mbox{SU(2)}}}
\def\Lphi{{\cal L}_{\mbox{$\Phi$}}}
\def\Lwz{{\cal L}_{\mbox{WZ}}}
\def\su2xsu2{{SU(2)\times SU(2)}}
\def\su3xsu3{{SU(3)\times SU(3)}}
\def\hr{{\hat r}}
\def\rhat{{\hat r}}
\def\fpi{{f_\pi}}
\def\fpis{{f_\pi^2}}
\def\Fpis{{F_\pi^2}}
\def\Fpi{{F_\pi}}
\def\vtau{{\vec \tau}}
\def\vrho{{\vec \rho}}
\def\trho{{\tilde \rho}}
\def\vT{{\vec T}}
\def\tG{{\tilde G}}
\def\tV{{\tilde V}}
\def\tomega{{\tilde \omega}}
\def\vOmega{{\vec \Omega}}
\def\vomega{{\vec \omega}}
\def\vbeta{{\vec \beta}}
\def\rvec{{\vec r}}
\def\vpi{{\vec\pi}}
\def\rhomn{\rho_{\mu\nu}}
\def\rhoMN{\rho^{\mu\nu}}
\def\omegamn{\omega_{\mu\nu}}
\def\omegaMN{\omega^{\mu\nu}}
\def\eps{{\epsilon^{\mu\nu\alpha\beta}}}
\def\Imua{{I_\mu^a}}
\def\be{\begin{equation}}
\def\ee{\end{equation}}
\def\ben{\begin{eqnarray}}
\def\een{\end{eqnarray}}
\def\C{{\cal C}}
\def\O{{\cal O}}
\def\J{{\cal J}}
\def\Gg{G_{\mbox{global}}}
\def\Hl{H_{\mbox{local}}}
\def\mzero{\mbox{\bf 0}}
\def\mone{\mbox{\bf 1}}
\def\Msol{M_{\mbox{sol}}}
\def\Mrot{M_{\mbox{rot}}}
\def\eg {{\it e.g.}}
\def\prl {Phys. Rev. Lett.}
\def\pl {Phys. Lett.}
\def\pr {Phys. Rev.}
\def\np {Nucl. Phys.}
\def\ie {{\it i.e.}}
\def\calF {{\cal F}_{\cal A}}
\def\calFp {{\cal F}_{\tilde{\cal A}}}
\def\pin {{\Pi_n (0)}}
\def\pint {{\Pi_n (X(t))i}}
\def\calA {\cal A}
\def\calAp {\tilde{\cal A}}
\newcommand{\ISC}[4]{\mid #1, #2; #3, #4 \rangle_I}
\newcommand{\Rs}[2]{\mid #1, #2 \rangle_R}
\newcommand{\Js}[2]{\mid #1, #2 \rangle_S}
\newcommand{\Jc}[2]{\mid #1, #2 \rangle_C}
\newcommand{\ptl}[1]{\mid #1 \rangle}
\newcommand{\fsq}[2]{\sqrt{\frac{#1}{#2}}}
\newcommand{\fun}[2]{\frac{#1}{#2 {{\cal J}}}}

\begin{document}
\begin{titlepage}
\begin{center}
\vskip 1.0cm
{\large\bf NUCLEAR AXIAL-CHARGE TRANSITIONS}\\
{\large\bf IN CHIRAL PERTURBATION THEORY}
\vskip 3cm
{\large Tae-Sun Park and Dong-Pil Min}\\
{\it Department of Physics and Center for Theoretical Physics}\\
{\it Seoul National University}\\
{\it Seoul 151-742, Korea}
\vskip 0.2cm
{and}
\vskip 0.2cm
{\large Mannque Rho}\\
{\it Service de Physique Th\'{e}orique, CEA Saclay}\\
{\it 91191 Gif-sur-Yvette, France}
\vskip 3cm
{\bf ABSTRACT}
\begin{quotation}
We develop a systematic chiral perturbation expansion for the calculation of
meson-exchange currents in nuclei and apply the formalism to nuclear axial
currents. We summarize the principal results of such a calculation to one loop
order on nuclear axial-charge transitions which provides a strong support to
the conjecture of ``chiral filter phenomenon" in nuclear medium.
The use of heavy baryon chiral perturbation theory enables us to obtain a
remarkably simple result valid next to the leading order in chiral counting.
The dominant role of a soft-pion exchange in axial-charge transitions in heavy
nuclei is confirmed. An important, {\it albeit} indirect, consequence of our
result on the empirically observed enhancement in axial-charge transitions in
heavy nuclei is pointed out.

\end{quotation}
\end{center}
\end{titlepage}

While meson-exchange currents in nuclei are fairly well understood
in low-energy and low-momentum regime
both experimentally and theoretically \cite{fm}, there remains the
issue of understanding them from the point of view of QCD. This is not
an academic issue since the purported high-energy electron machines in
construction or in project are to probe nuclei for signals of direct
QCD degrees of freedom in the {\it deviation} from exchange-current effects
and for this, a contact with QCD at lower energy
and highly nonperturbative regime will be clearly needed. At present the only
applicable method to address this issue in nonperturbative regime is
chiral perturbation theory (ChPT) \cite{gl}.
Chiral perturbation theory has had much success in Goldstone boson ($\pi$, $K$)
interactions\cite{gl}, making nonperturbative QCD accessible to laboratory
phenomena. However incorporating baryons in the scheme has proven
to be quite difficult, the main reason being that the standard
power counting used in ChPT does not apply when baryons are involved.
Even for the elementary $\pi$-N interaction, the
calculation becomes horrendously complicated, requiring approximations
that are hard to control. Some progress has recently been made in
describing by chiral perturbation expansion such processes as $\pi$-N
scattering \cite{gss}, threshold pion production from nucleon \cite{bkm}
and nucleon's electromagnetic polarizabilities \cite{bkmp}. Nonetheless
a systematic chiral perturbation calculation
of many-body nuclear properties remains a formidable, if
not hopeless, task at the present stage of development. This technical
difficulty explains in part the paucity up to date
of work of this nature in the literature. Recently, however,
this task was greatly facilitated by the formulation of heavy-baryon
chiral perturbation theory (which we shall call HBChPT in short)
by Weinberg \cite{wein90}, Jenkins and Manohar
\cite{jm91} and others \cite{pw91} along the line developed for heavy-quark
effective theory \cite{georgi}.

The purpose of this letter is to summarize the principal results of
a complete chiral perturbation calculation of the meson-exchange axial-charge
operator to one-loop order that establishes rigorously
the ``chiral filter phenomenon" in nuclei conjectured
a long time ago \cite{kdr} and given a partial justification recently
by one of the authors \cite{mr91} which states that {\it whenever kinematically
unsuppressed, soft-pion exchanges should dominate in electroweak processes
in nuclei}. As discussed in \cite{kdr}, the chiral filter phenomenon occurs in
nuclear axial-charge transitions and in nuclear M1 transitions (such as
threshold radiative np capture). In this paper, we will focus on the former.
The latter will be discussed in a separate paper in preparation \cite{pmr}.
Towner \cite{towner} and Riska {\it et al} \cite{riska} have recently
addressed a similar issue within, however, a phenomenological framework.
We shall make comparison with their results at the end of the paper.

As in \cite{wein90,mr91}, we shall take the most general chiral Lagrangian
consisting of nucleons and pions with all other degrees of freedom integrated
out. (One can alternatively take a Lagrangian that also contains vector mesons
and $\Delta$'s. We have satisfied ourselves that these elements do not alter
our result significantly. This matter will be discussed in detail in a
later publication.) The key observation made by the authors in
\cite{wein90,jm91} is that the standard derivative expansion successful in
low energy pion dynamics breaks down when baryons are involved, for the
reason that time derivatives on the baryon field, typically of the order
$m_B\sim 1$ GeV, are not small on the chiral scale set by
$\Lambda_\chi \sim 1$ GeV.
The heavy-baryon formalism HBChPT circumvents this
difficulty. To set up the formalism, one first redefines
the velocity-dependent baryon field \cite{georgi}
$B_v (x)=e^{i m_B \gamma\cdot v\, v\cdot x} N(x)$
constrained to $\not\!{v} B_v=B_v$ where $v_\mu$ is the four-velocity with
$v^2=1$ and $m_B$ the mass of the baryon.
This redefinition of the baryon field renders the relevant momentum scale
of the baryon involved to be of the same
order as the pion momentum. Consequently, HBChPT allows a chiral expansion
in power of $(\partial / \Lambda_\chi)$ and $(\partial / m_B)$
corresponding to the pion momentum scaled by $\Lambda_\chi$ and the
residual baryon momentum (when $m_B v_\mu$ is removed) scaled by $m_B$,
respectively.
Expressed in terms of the $B$ field, the
chiral Lagrangian of pions and nucleons takes the form\footnote{
We can equally well rewrite this form
in terms of the Sugawara field $U=e^{2i\pi/ F_\pi}$
frequently used in the literature \cite{jm91,pw91}.}
\ben
\L&=& \frac{1}{2}D_\mu\vec{\pi}\cdot D^\mu \vec{\pi} -\frac{1}{2} D^{-1}
m_\pi^2
\vec{\pi}^2
+ \bar{B_v}\left[iv\cdot {\cal D}
+2\frac{g_A}{F_\pi} S_v^\mu \vec{\tau}
\cdot D_\mu \vec{\pi}\right] B_v\nonumber\\
&-& \frac{1}{2}\sum_\alpha \left(\bar{B}_v \Gamma_a B_v\right)^2 +\cdots
\label{chilag}
\een
where ${\cal D}_\mu=\del_\mu +i \frac{1}{F_\pi^2} \vec{\tau}\cdot
\vec{\pi}\times D_\mu \vec{\pi}$ and
$D_\mu=(1+\vec{\pi}^2/F_\pi^2)^{-1}\del_\mu\equiv D^{-1}\del_\mu$ are the
``covariant derivatives" and $S_v^\mu\equiv \frac{1}{4}\gamma_5
[\not\!\!{v},\gamma^\mu]$ is the spin operator satisfying $v_\mu S_v^\mu=0$.
The $ \Gamma_a$ in the quartic fermion term of Eq.(\ref{chilag}) stands for
all possible terms containing no derivatives allowed by Lorentz invariance
and chiral symmetry and the ellipsis for higher-derivative and quark mass
terms allowed by symmetries. In our case, higher derivatives acting on
pion fields and quark mass terms play no essential role,
so will be ignored. We will later make a specific
reference to higher derivative terms
involving baryons in connection with our renormalization procedure.
The merit of this Lagrangian is that it provides a consistent chiral
expansion for {\em both mesons and baryons} in terms of
derivatives and quark masses \cite{wein90,jm91}.\footnote{The HBChPT as
formulated is essentially equivalent to making static
approximation on baryons fields, with the baryon velocity effectively conserved
and with typical off-shell momentum of the baryon counted as of the same order
as the pion momentum.} Baryon momentum-dependent terms appear as higher
order interaction terms in the chiral counting. The pair terms are suppressed
at the leading order, appearing as ``$1/m$" corrections subsumed in the
ellipsis in (\ref{chilag}).

Our task is to compute higher-order corrections in the chiral expansion
parameter $Q$ -- where $Q$ is the momentum or energy scale parameter probed by
an external field, assumed to be small compared with the chiral scale
$\Lambda_{\chi}$ -- to the leading soft-pion amplitude, denoted
${\cal M}_{soft}$, that contributes in two-body {\it effective} currents
responding to a slowly varying electroweak field $J_\mu$,
\ben
{\cal M}={\cal M}_{soft} (1+\delta+ O(Q^n)),\ \ \ \
\delta\sim O(Q^2),\ \ \ \ n\geq 3.\label{amp}
\een
There are no corrections of $O(Q)$ to $\delta$ for the same reason that they
are
absent in nucleon-nucleon potentials \cite{wein90}.
This chiral counting was established in \cite{mr91}
for both electromagnetic and axial exchange currents.
In what follows we will focus on the axial-charge operator as it is
currently receiving considerable phenomenological attention \cite{warburton}.
We will use the axial current ${\cal A}^\mu_a$ obtained from the Lagrangian
(\ref{chilag}) by Noether construction, {\ie}, ${\cal A}^\mu_a=
\sum_{\phi=B_v,\pi_b}\frac{\del{\L}}{\del(\del_\mu \phi)}[X_a,\phi]$.

As emphasized recently by Weinberg \cite{wein90,wein92}, one should
restrict the use of chiral perturbation strictly to {\it irreducible} graphs
that are free of small energy denominators responsible for binding.
Exchange currents belong to this class of graphs.
Now corrections to the soft-pion result can come from two sources: from
contact four-fermion interactions involving derivatives and from one-loop
graphs.
We shall argue below that there should be no contribution from the former to
the
order we are considering. As for the latter, there are three classes of
one-loop graphs to be calculated:
The first (denoted as A) consists of one-pion-exchange current with one-loop
renormalization of either the internal pion line or an external nucleon line
or the $\pi NN$ vertex or the $J_\mu\pi NN$ vertex; the second (denoted as B)
consists of one-loop two-pion-exchange current with or without nucleon
intermediate states; and the third (denoted as C) involves the four-fermion
contact interaction $\Gamma$ in (\ref{chilag}).
Most of the diagrams in A can be absorbed into the standard perturbative
renormalizations of the pion mass, the nucleon mass and wavefunction and
of the constants $F_\pi$ and $g_A$. Since to one loop in the HBChPT there are
no momentum-dependent corrections to these quantities, they are trivial
and do not directly figure in our result.
The only nontrivial contribution in this class comes from the $J_\mu\pi NN$
vertex, the relevant diagrams of which are given in Fig.A($a$-$n$). The class-B
graphs are given in Fig.B($a$-$h$) and the class-C in Fig.C($a$-$b$).

At first sight, the number of graphs may look daunting but in the HBChPT with
the Lagrangian (\ref{chilag}), many of these graphs
do not contribute to the chiral order we are
interested in. Some graphs vanish identically due to isospin symmetry
({\it viz}, Fig.B$e$). Other graphs such as Figs.C$a$, A$i$, A$j$, A$k$,
A$l$, A$m$ and A$n$ vanish because they are proportional to
$v\cdot S_v$ which is zero.
The graphs C$b$, A$g$, A$h$, B$f$, B$g$ and B$h$ are proportional to
$S_v^\mu$, hence do not contribute to the time component since
$S_v^0\sim O(Q/m_B)$. Thus we are left with only four graphs B($a$, $b$,
$c$, $d$) in the classes B and C and six graphs A($a$-$f$) in the class A.
We should note that
while contact four-fermion interactions are very important in nuclear forces
\cite{wein90,pw91}, they are suppressed not only at the leading
chiral order as shown in \cite{mr91} but also at higher orders. This feature
is manifest in Weinberg's form of nonlinear chiral Lagrangian,
the consequence of which constitutes one of the key elements in this work.
This is in stark contrast with nuclear forces where numerous counter
terms that are not readily available from experiments make systematic loop
corrections highly problematic \cite{wein92}.

We shall take the momentum carried by the current to be zero and evaluate
the two-body amplitudes corresponding to the exchange axial-charge operator.
The HBChPT renders the loop integrals doable analytically.
We note that even after the usual (wavefunction, mass and coupling constant)
renormalization, there are additional (logarithmic) divergences left over
({\ie}, $L$ defined below). This is expected as we are dealing with a
nonrenormalizable theory. This however poses no difficulty in our case.
As is the standard practice \cite{gl}, these divergences can be absorbed
in the coefficients of the counter terms that are next order in chiral
expansion. The counter terms that absorb {\it all} the
one-loop divergences in our calculation -- that are subsumed in the ellipsis
in (\ref{chilag}) -- are (in Weinberg representation) of
the following form involving higher derivatives:
\ben
{\cal L}_{ct} &=&
-i\frac{d_2}{F_\pi^2} \,\bar{B}_v \left[ {\cal D}^\mu, \left[v\cdot {\cal D},
{\cal D}_\mu\right]\right] B_v
\nonumber \\
&-& \frac{4 g_A}{F_\pi^5} v^\mu D_\mu \vec{\pi} \cdot
\left\{ d_4^{(1)}(\bar{B}_v \vec{\tau}{\cal D}_\nu B_v) \times
(\bar{B}_v\vec{\tau} S^\nu_v B_v) \right.
\nonumber \\
&&+\ \left. i \,d_4^{(2)}
\left( (\bar{B}_v \left[S^\alpha,
S^\beta\right] \vec{\tau} {\cal D}_\beta B_v)\,(\bar{B}_v S_\alpha B_v)
+(\bar{B}_v \left[S^\alpha,
S^\beta\right] {\cal D}_\beta B_v)\,(\bar{B}_v \vec{\tau}S_\alpha B_v)
\right)\right.
\nonumber \\
&&\left. +\ h.c. \right\}
\label{ctl}
\een
with
$d_2 = \kappa_2 + \frac{1}{24\pi^2} (1+5 g_A^2) \eta$,
$d_4^{(1)} = \kappa_4^{(1)} + \frac{1}{16\pi^2} (3 g_A^2 -2) \eta$ and
$d_4^{(2)} = \kappa_4^{(2)} - \frac{1}{2\pi^2} g_A^2 \eta$ where
$\eta= \frac{2}{4-d}+ \Gamma'(1)+{\rm ln}(4\pi)  - {\rm ln}(m_\pi^2)$,
$\Gamma'(1) \simeq - 0.577215$
and $\kappa$'s are finite constant counter terms.\footnote{
$\eta$ and $d$'s are singular and contain logarithms of mass.
Here we briefly sketch our renormalization prescription.
In calculating loop graphs, we use dimensional regularization.
We encounter singular quantities in the form of $\eta$ as given above.
To remove them, we write a counter-term Lagrangian which is formally
of the same chiral order as the one-loop graphs we compute.
We adjust the coefficients of this counter-term
so as to obtain a regular expression. The constants $d_i$ so introduced
contain two parts: one is proportional to $\eta$ and removes divergences
and the other, the finite constants $\kappa_i$ which are to be
determined (in principle) from experiments. The renormalization
will be done at on-shell point for the nucleon and at zero-momentum
for the pion. To one loop, the pion mass $m_\pi$ and the decay constant
$F_\pi$ are independent of the renormalization point and can be
taken from experiments.}
Now performing the calculation with dimensional regularization, we find
\ben
\delta(1\pi) &=& \frac{Q^2}{F_\pi^2}\left[ \kappa_2 +
\frac{1+ 3 g_A^2}{8 \pi^2} K_0(Q^2) - \frac{1+2 g_A^2}{2\pi^2} K_2(Q^2)
\right],
\nonumber\\
\delta(2\pi) &=&
\frac{Q^2 + m_\pi^2}{F_\pi^2} \left[ \kappa_4^{(1)}
+ \kappa_4^{(2)}\, \xi
+ \frac{3g_A^2-2-8 g_A^2\, \xi}{16\pi^2} K_0(Q^2)
+ \frac{g_A^2}{8\pi^2} K_1(Q^2)\right]
\label{afinal}
\een
where $q_\mu$ the four-momentum transferred from nucleon `1' to nucleon `2'
carried by pions,
$K_0 (Q^2)= -2+\sigma y$,
$K_1 (Q^2)= 1-\frac{\sigma^2-1}{2\sigma} y$ and
$K_2 (Q^2) = - \frac{4}{9} + \frac{\sigma^2}{6}
+ \frac{\sigma(3-\sigma^2)}{12}y$
with $y\equiv {\rm ln}\frac{\sigma+1}{\sigma-1}$,
$\sigma\equiv \sqrt{\frac{4m_\pi^2+Q^2}{Q^2}}$.
Note that all the constants appearing here are
physical ones which should be identified with experimental values.
We have replaced at appropriate places
the momentum $q_\mu^2$ by the expansion scale parameter $-Q^2$.
The ratio of the spin-isospin matrix elements
$\xi=\frac{\langle \,i (\vec{\tau}_1+ \vec{\tau}_2)\, \vec{q} \cdot
\vec{\sigma}_1\times\vec{\sigma}_2\, \rangle}{
\langle \, i \vec{\tau}_1\times \vec{\tau}_2 \,\vec{q} \cdot
(\vec{\sigma}_1+\vec{\sigma}_2)\,\rangle}$
figuring in the two-pion contribution is introduced for convenience.
This ratio can be estimated in various nuclear models:
In simple jj shell model, in Wigner supermultiplet model
as well as in Fermi gas model of the nucleus,
the ratio comes out to be 1\footnote{
We would like to thank K. Kubodera for an invaluable help on this ratio.}
so we shall set $\xi=1$ in the numerical estimates made below.
As defined in (\ref{afinal}), the quantity $\delta(1\pi)$ comes from
$J_\mu \pi NN$ vertex renormalization
for which only the six graphs Figs.$Aa$-$Af$ survive to the
chiral order we are calculating and $\delta(2\pi)$ from the surviving two-pion
exchange graphs.

A close inspection shows that the terms
involving $d_4$ in (\ref{ctl}) {\it cannot}
arise from single vector-meson exchanges or other excitations that are lower
than the chiral scale $\Lambda_\chi\sim 1$ GeV. Furthermore a
constant term proportional to $\kappa_4^{(1,2)}$ (plus other counter terms
that are in principle present even though no regularization is required)
implies a zero-range interaction depicting the exchange of very massive
degrees of freedom. Since we are to apply chiral perturbation expansion
to only {\it irreducible graphs} while all reducible graphs are to be taken
into account in calculating nuclear wave functions from a Schr\"{o}dinger
equation (or a relativistic generalization thereof) with a potential
defined with the irreducible graphs and consequently the nuclear wave
functions so obtained must contain short-range correlations,
the consistency with the scheme requires that
when embedded in nuclear matter, such a contact term be suppressed by
nucleon-nucleon correlations. This invites us to drop the constant
terms (or $\delta$ function terms in coordinate space)
\footnote{Purists might object to this procedure by arguing that
one has to calculate {\it both} nuclear forces and current matrix elements
to the same order of chiral expansion. This we believe is {\it not} the
right way of using chiral perturbation theory in nuclei. In fact, it is
not a fruitful way of doing physics as a little thought would reveal
that such a so-called ``consistent" approach is doomed to fail.
This ``failure" should however not be construed as a failure of ChPT in
nuclear physics as some people seem to argue.}.
The remaining constant $\kappa_2$, which figures in the calculation of
$A_\mu \pi N N$ vertex, can be fixed by the isovector
charge radius of the nucleon,
\be
\kappa_2 = - \frac{1}{6} F_\pi^2 \langle r^2 \rangle_1^V \simeq - 0.0856.
\ee
This results because
the $A^\mu \pi N N$ vertex is related to the isovector Dirac form factor
of the nucleon, $F_1^V(t)$. This can be understood also by current algebra
or in terms of vector-meson exchange \cite{pmr}.

%

One can have a rough idea of how large the chiral loop corrections can be
by taking, in eq.(\ref{afinal}), $Q\sim m_\pi$ as befits the scale involved
in the chiral expansion. It comes out to be
\be
|\delta (Q\approx m_\pi)| \leq 0.05.
\ee
To make a more quantitative estimate in nuclear matter, we have to
go to coordinate space. It is in this space that short-range correlations
mentioned above are most straightforwardly taken into account.\footnote{In
momentum space, this procedure roughly corresponds to subtracting constant
terms from the expressions of (\ref{afinal}). This is analogous to the
constant subtraction one does to incorporate the Lorentz-Lorenz effect
in $\pi$-nucleus scattering. However in the present case, because of
non-analytic terms, such simple prescriptions are not reliable.
We have no choice but to go to coordinate space.}
Let us write the two-body axial-charge operator as
\be
{\cal M} = (1+\delta_{soft}) {\cal M}_{soft} + {\cal M}_{loop}
\label{Mcoordi}\ee
where
\be
\delta_{soft} = - \kappa_2 \frac{m_\pi^2}{F_\pi^2}
+ \frac{m_\pi^2}{4\pi^2 F_\pi^2}
\left[\frac{1+3 g_A^2}{2} \left(2 -\frac{\pi}{\sqrt{3}}\right)
-(2 + 4 g_A^2) \left(\frac{13}{9} - \frac{\pi\sqrt{3}}{4}\right)\right]
\simeq 0.0455.
\ee
In (\ref{Mcoordi}), we have separated out
the long-range ($\O(Q^2)$) contribution, denoted $\delta_{soft}$,
from one-loop corrections to the one-soft-pion exchange.
The remainder is shorter-ranged and hence combined with two-pion-exchange
contribution into ${\cal M}_{loop}$, representing the nontrivial part of
one-loop corrections. The expression
for ${\cal M}_{loop}$
is rather involved and so will not be written down explicitly in this paper.
See \cite{pmr} for details. In Fig. 2 is plotted the quantity $4\pi r^2
{\cal M} (r)$ where $r=|\vec{r}_1 -\vec{r}_2|$. The axial-charge matrix
element in nuclear medium calculated in fermi-gas model with a hard-core
cut-off of $d=0.4 - 0.7$ fm is given in
Fig. 3 as a function of matter density. Two key features to note in these
results are: First, the loop corrections are very small for
$r \geq 0.6$ fm compared with the
soft-pion term consistent with the chiral filter argument and second
the density dependence of the loop correction is weak.
More precisely, for the reasonable hard-core cut-off of $d=0.5$ fm,
the ratio of loop correction over soft-pion
$R\approx 0.067$ for $\rho=0.5 \rho_0$ and $\approx 0.089$
for $\rho=\rho_0$. Even at nuclear matter density, the loop correction
represents less than 10\% of the soft pion result.

To summarize, we have shown that the loop corrections to the
soft-pion exchange axial-charge operator can be easily calculated
in the HBChPT that provides a consistent chiral expansion. In this formalism,
nucleon-antinucleon pairs are suppressed to next to the leading order.
They can only contribute at higher chiral orders.
Here we focused on the axial-charge operator but the same calculation
can be done with no greater difficulty for the space component of
the vector current, {\it e.g.}, the M1 operator relevant for
the electrodisintegration of the deuteron where the soft-pion effect is
even more spectacular \cite{pmr}.
An important outcome of our calculation to
one loop is that the chiral filter mechanism
is robust, the soft-pion term playing a predominant role with
one-loop chiral corrections remaining in the noise at small momentum transfers.
It is hopeful that it will survive higher loop corrections.
It is pleasing that our result eq.(\ref{Mcoordi}) is totally free of unknown
parameters inherited from more massive degrees of freedom once it is accepted
that short-range correlations are operative in the transition matrix elements.

We should point out one important, though indirect, consequence of our
result. Since the suppression of higher-order chiral corrections established
in this paper is likely to persist independently of the environment into which
the operator is embedded, it is natural to conclude that higher-order chiral
corrections {\it cannot} generate the apparently significant
density dependence of the meson-current enhancement in heavier nuclei observed
in nature \cite{warburton}.
The present calculation strongly suggests that the origin of the apparent
density dependence of the axial charge transitions observed in Nature
lies outside of higher-order chiral corrections, thus pointing to the
possibility that the basic mechanism for the enhanced axial charge in heavy
nuclei is indeed, as proposed in \cite{kr}, the scaling of hadron masses
and quark condensate in dense medium, which is an intrinsic vacuum property
\cite{br91}. Since the argument is based simultaneously
on chiral symmetry and scale anomaly of QCD, it
is perfectly consistent with the chiral expansion. We note that an empirical
support for the scaling notion comes from a recent experiment on
$^{10}B (\vec{p},\vec{p}^\prime)$ at 200 MeV \cite{pp}.

This conclusion raises the question as to how the phenomenological
Lagrangian approach of \cite{towner,riska} can be understood in terms
of the result we have obtained here. In particular, since the HBChPT
relies on the suppression of pairs, it is natural to ask whether or if so,
how the pair terms
involving heavy mesons of \cite{towner,riska} can be interpreted. The
answer to this question that we propose is that at least the main pair
term that figures in \cite{towner,riska}, namely the $\sigma$ exchange,
corresponds to replacing the mass of the nucleon $m_B$ by the medium
quantity $m_B^*$ in the sense suggested in \cite{br91},
with the vector meson mediated
pair terms suitably suppressed by short-range correlations. A similar
suggestion has been made in \cite{wt}. This means that
while two-body pair term in \cite{towner,riska} renormalizes the mass in the
single-particle operator, three-body operators involving pairs will be needed
to renormalize the two-body operator in the way it figures in \cite{kr}.
It would be interesting to verify this by an explicit calculation with
phenomenological Lagrangians.
\subsubsection*{Acknowledgments}

We are grateful for useful discussions with Gerry Brown and Kuniharu
Kubodera. We would also like to acknowledge helpful comments on
our work from Ulf Meissner.
This work was initiated while one of the authors (M.R.) was visiting
the Center for Theoretical Physics, Seoul National University. He wishes to
thank the members of the Center for their hospitality. It was supported
in part by the KOSEF through the Center for Theoretical Physics.

\pagebreak

\newpage
\centerline{\bf FIGURE CAPTIONS}
\vskip 1cm
\begin{quotation}
\noindent{\bf Fig. 1}
\noindent
\begin{itemize}
\item A: The class-A one-loop graphs that renormalize the $J_\mu\pi NN$ vertex
contributing to one-pion exchange axial-charge
operator. The cross represents the axial current, the solid
line the nucleon and the dotted line the pion. Only the graphs
$(a)$, $(b)$, $(c)$, $(d)$, $(e)$ and $(f)$ survive to contribute.
\item B: The class-B one-loop graphs $(a)$-$(h)$ for
two-pion-exchange axial-charge operator. Only the graphs $a$, $b$, $c$ and $d$
survive.
\item C: The class-C one-loop
graphs $(a)$-$(b)$ involving four-fermion interaction.
Both graphs do not contribute to the chiral order considered.
\end{itemize}
\vskip 0.5 cm
\noindent{\bf Fig. 2}
\noindent
\begin{itemize}
\item Two-body axial-charge operator as a function of the separation distance
obtained by putting
$\tauvec_1\times\tauvec_2 \,(\sigma_1+\sigma_2)\cdot \hat{\bf r}
=(\tauvec_1+\tauvec_2)\, \sigma_1\times\sigma_2\cdot \hat{\bf r}=1$.
Solid line is $4\pi r^2 (1+\delta_{soft}) {\cal M}_{soft}$ and dotted line is
$4\pi r^2 {\cal M}_{loop}$.
\end{itemize}
\vskip 0.5cm
\noindent{\bf Fig. 3}
\begin{itemize}
\item The ratio of the matrix elements $R=\langle {\cal M}_{loop}\rangle/
\langle (1+\delta_{soft}){\cal M}_{soft} \rangle$ as a function of
$\rho/\rho_0$ for various hard-core cut-off $d$ in the cut-off function
$\theta (r-d)$.
\end{itemize}

\end{quotation}


\end{document}